\documentclass[12pt]{article}
\usepackage{a4}
\usepackage{graphicx}
\usepackage{epsfig}

\def\lsim{\;
\raise0.3ex\hbox{$<$\kern-0.75em\raise-1.1ex\hbox{$\sim$}}\;
}

\def\gsim{\;
\raise0.3ex\hbox{$>$\kern-0.75em\raise-1.1ex\hbox{$\sim$}}\;
}

\begin{document}
%\date{\today}

\renewcommand{\thefootnote}{\fnsymbol{footnote}}

\title{{\bf \large Note on finite temperature sum rules for vector
and axial-vector spectral functions
}\thanks{Work supported in part by BMBF and GSI.}
}

\author{E. Marco$^a$\footnotemark[7]~, R. Hofmann$^b$ ~and W. Weise$^{a,c}$\\
\\
$^a$Physik-Department,\\
Technische Universit\"at M\"unchen,\\
D-85747 Garching, Germany\\
\\
$^b$Max-Planck-Institut f\"ur Physik\\
(Werner-Heisenberg-Institut)\\
D-80805 M\"unchen, Germany\\
\\
$^c$ECT$^{*}$, I-38050 Villazzano (Trento), Italy\\
}
\maketitle

\footnotetext[7]{Fellow of the A.v.Humboldt Foundation.}

\vspace{-15cm}
\begin{flushright}
{\bf TUM/T39-01-23\\ 
 MPI-PHT 2001-38}
\end{flushright}
\vspace{13cm}

\begin{abstract}
An updated analysis of vector and axial-vector spectral functions
is presented. The resonant contributions to the spectral integrals 
are shown to be expressible as multiples of $4 \pi^2 f_{\pi}^2$,
encoding the scale of spontaneous chiral symmetry breaking in QCD. 
Up to order $T^2$ this behavior carries over 
to the case of finite temperature. 
\end{abstract}

\newpage

Quantum Chromodynamics (QCD) has an approximate chiral
symmetry which is spontaneously broken. This Nambu-Goldstone
realization of chiral symmetry manifests
itself through the presence of a quark condensate 
$\langle \bar{q} q \rangle$, pseudoscalar Goldstone bosons and a
non-vanishing pion decay constant, $f_{\pi}\simeq 92.4$ MeV, which
sets the characteristic chiral ``gap'' scale, $4 \pi f_{\pi}\sim 1$ GeV.
The symmetry breaking pattern is also evident in the difference between
the spectral functions of vector $(J^{PC} = 1^{- -})$ and
axial-vector $(J^{PC} = 1^{+ +})$ channels. These spectral
functions would be identical if chiral symmetry were realized in
the (unbroken) Wigner-Weyl mode.

At high temperatures, $T> T_c \simeq 0.2$ GeV, chiral
symmetry is expected to be restored. The vector and axial-vector
channels should become degenerate as $T$ approaches $T_c$. The
leading temperature dependence of the corresponding current correlators
has been derived in refs.\ \cite{DeyEle,EleIof}. The result is that
vector and axial-vector modes mix at finite $T$ as a consequence
of their couplings to the pionic heat bath
(see also ref.\ \cite{Urb}), but that their
masses remain unchanged to leading order, $O(T^2)$.

The $T=0$ vector $(V)$ and axial-vector $(A)$ spectral functions have recently
been determined with high accuracy up to invariant masses 
$\sqrt{s} = m_{\tau} \simeq 1.78$ GeV (the tau lepton mass) by
the ALEPH \cite{ALEPH} and OPAL \cite{OPAL} collaborations. These data
serve as a reliable basis for all sorts of sum rule analysis (see e.g.\
ref.\ \cite{IofZya}), at least for the lowest moments of the $V$ and $A$
spectral distributions.

The purpose of our present note is twofold. First we follow up on a previous
study \cite{FESR}, using finite energy sum rules (FESR), in which
the chiral ``gap'' $4 \pi f_{\pi}$ was suggested as the dominant scale
governing the spectral sum rules (see also ref.\ \cite{GolPer}). We
give an interpretation of how this scale is encoded in the pattern
of the measured $V$ and $A$ spectral distributions and (as a reminder)
in the Weinberg $(V-A)$ sum rules. Second we study the case of finite
temperature $T$. The change of the chiral ``gap''
with temperature is determined  by the $T$ dependence
of the pion decay constant (in fact, the one related to
the time component of the axial current), as given in ref.\ 
\cite{GasLeuT}. We investigate spectral sum rules for the $T$ dependent $V$
and $A$ correlators and examine their consistency with the
Eletsky-Ioffe result \cite{DeyEle,EleIof} to order $T^2$.

We begin with the vector and axial-vector correlators at zero
temperature. Let $J^{\mu}$ be either one of the currents
$V^{\mu} = \bar{u}\gamma^{\mu} d$ or 
$A^{\mu} = \bar{u}\gamma^{\mu}\gamma_5 d$ {
interpolating to the negatively charged member of the
corresponding mesonic isotriplet.}
The correlators

\begin{equation} \label{eq:correlatorT0}
\Pi^{\mu \nu}_{J} (q) = i \int d^4x\, e^{i q x}
\langle 0| {\cal T} [J^{\mu}(x) J^{\nu}(0) ]
| 0 \rangle
\end{equation}
can be decomposed as

\begin{equation} \label{eq:corr_escalar}
\Pi^{\mu \nu}(q) = (q^{\mu}q^{\nu} - q^2 g^{\mu \nu})
\Pi^{(1)} (q^2) + 
q^{\mu}q^{\nu} \Pi^{(0)}(q^2)\, ,
\end{equation}
in terms of their spin-0 and spin-1 parts, $\Pi^{(0)}$ and $\Pi^{(1)}$,
respectively. The spin-0 part is relevant only for the axial-vector
current where it represents the induced pseudoscalar (pion pole)
contribution. The $V$ and $A$ spectral functions are defined as

\begin{equation}
v_1(q^2) = 2 \pi \mbox{Im} \Pi^{(1)}_V (q^2)\, ,
\end{equation}

\begin{equation}
a_{0,1}(q^2) = 2 \pi \mbox{Im} \Pi^{(0,1)}_A (q^2)\, , \ \ \ \ {(q^2\equiv s\ge0)\ .}
\end{equation}
The pion pole term is

\begin{equation}
a_0(q^2) = 4 \pi^2 f^2_{\pi} \delta(q^2-m_{\pi}^2)\, .
\end{equation}
Given the analytical properties of $\Pi^{(1)}_V (s)$ and
$\Pi^{(0)}_A (s)+\Pi^{(1)}_A (s)$ in the complex $s$-plane, we
now write FESR relations for their spectral functions
\cite{FESR,Chetyrkin}. Assume that the {spectral continuum is described in 
terms of perturbation theory at $s \ge s_c$ and that 
only the first few mass dimensions are relevant in the operator product 
expansion (OPE) of the current correlators evaluated at $|q^2|=s_c$}.
Choose a closed path which surrounds the cut along the real $s$-axis and
joins a circle of radius $s_c$. The Cauchy integral around this
closed path includes the integration along the circle which {can be estimated 
using the OPE.\footnote{{If $\sqrt{s_c}\gg\lambda^{-1}$, where $\lambda$ 
denotes a typical correlation length characterizing gauge invariant
two point functions, which generalizes the local condensates, then there 
are no oscillations in the Minkowski-like and no exponential
suppression in the euclidean-like domains 
\cite{Hofmann1}.}}} {By means of the optical theorem 
the integral along the positive, real axis is evaluated 
using experimental cross sections directly or by 
considering a model motivated by observation.} For details see ref.\
\cite{FESR}. {As far as quark mass corrections in the OPE (mass 
perturbation in the perturbative part and the term involving the quark
condensate) are concerned, we neglect them,
which is justified by their
numerical smallness\footnote{At $T>0$ there are 
to order $T^2$ corrections to the gluon condensate and contributions
from O(3) invariant operators proportional to $m_\pi^2$ \cite{Hatsuda}
which hence vanish in the chiral limit. 
However, if the {\sl vacuum state} is allowed to be affected by
the heat bath (not addressed in this work) then the 
gluon condensate does exhibit a $T$ dependence \cite{Hofmann2}.}. 

A compendium of the OPE for vector and axial-vector
correlators can be found in ref.\ \cite{BraNar}. As a result of this procedure
one finds the following sum rules for the lowest moments of $v_1$
and $a_0+a_1$:

\begin{equation} \label{eq:0th}
\int_0^{s_c} ds\, v_1(s)= \int_0^{s_c} ds\, [a_0(s) +a_1(s)]=
\frac{s_c}{2} (1+\delta_0)\, ,
\end{equation}
with the pion pole contribution

\begin{equation} \label{eq:pipole}
\int_0^{s_c} ds\, a_0(s) = 4 \pi^2 f^2_{\pi}\, ,
\end{equation}
and

\begin{equation} \label{eq:1st}
\int_0^{s_c} ds\, s \,v_1(s)= \int_0^{s_c} ds\, s \,a_1(s)= 
\frac{s_c^2}{4}(1+\delta_1) - \frac{\pi^2}{6} 
\langle \frac{\alpha_s}{\pi} G^2\rangle\, .
\end{equation}
The { right-hand sides} of eqs.\ (\ref{eq:0th},\ref{eq:1st}) 
include the {radiative corrections $\delta_{0,1}$
computed in perturbative QCD.} Their explicit expressions up to
order $\alpha_s^3$ are given in ref.\ \cite{FESR}. The first
moments (\ref{eq:1st}) introduce the dimension 4 gluon condensate
{$\langle (\alpha_s/\pi) G^2\rangle_{\mu\sim\sqrt{s_c}} 
\simeq (0.36\ \mbox{GeV})^4$}. Higher
moments involve condensates of correspondingly higher dimensions
and will not be considered here. 
The Weinberg sum rules (WSR) \cite{WSR} follow immediately:

\begin{equation} \label{eq:WSR1}
\int_0^{s_c} ds\, [v_1(s)-a_1(s)] = 4 \pi^2 f_{\pi}^2\, ,
\end{equation}
\begin{equation} \label{eq:WSR2}
\int_0^{s_c} ds\, s [v_1(s)-a_1(s)] = 0\, 
\end{equation}
where the limit $s_c \rightarrow \infty$ can be taken since the
high-energy continuum parts of $v_1$ and $a_1$ are identical. In
practice this asymptotic behaviour is reached at $s_c \simeq$~5~GeV$^2$.

Before turning to the temperature dependent spectral functions,
let us introduce a model of $v_1$ and $a_1$ which explicitly involves the
scale of spontaneous chiral symmetry breaking, $4 \pi f_{\pi}$. We
recall that in refs.\,\cite{FESR,GolPer} an appropriate large-$N_c$
representation of the vector spectrum was shown to be

\begin{equation}
v_1(s) = 8 \pi^2 f_{\pi}^2 \delta(s - m_{\rho}^2)
+ \frac{1}{2}\theta(s-s_0)\, ,
\end{equation}
with the $\rho$ meson mass $m_{\rho}$ and the continuum threshold
$s_0$ both expressed in terms of the chiral scale as
$\sqrt{2} m_{\rho} = \sqrt{s_0}=4 \pi f_{\pi}$. The contribution
of the $\rho$ meson to the WSR (\ref{eq:WSR1}) is then $8 \pi^2 f_{\pi}^2$,
twice that of the pion pole (\ref{eq:pipole}). This is 
what is seen in the data \cite{ALEPH,OPAL} when taking the WSR
integral up to $s\simeq 1$ GeV$^2$, covering the $\rho$ resonance.
The phenomenological bookkeeping seems to follow a pattern
in which the $n$-pion sectors of the spectrum each contribute
$n$ units of $4 \pi^2 f_{\pi}^2$ to the spectral integral, with
the $\rho$ meson collecting the strength in the $n=2$ sector,
for example. This conjecture suggests a parametrization

\begin{equation} \label{eq:v1}
v_1(s) = 4 \pi^2 f_{\pi}^2 [2d_{\rho}(s)+4d_{\rho'}(s)] + \mbox{continuum}\, ,
\end{equation}
\begin{equation} \label{eq:a0a1}
a_0(s)+a_1(s) = 4 \pi^2 f_{\pi}^2 [d_{\pi}(s)+3d_{a_1}(s)]
+ \mbox{continuum}\, ,
\end{equation}
where the distributions $d_n(s)$ are normalized as $\int ds\, d_n(s) =1$.
In the ``zero width'' (large $N_c$) limit we {have} 
$d_n(s) = \delta (s-m_n^2)$ with $m_1 = m_{\pi}$,
$m_2=m_{\rho}=\sqrt{2} \cdot 2 \pi f_{\pi}$, $m_3=m_{a_1}=4\pi f_{\pi}$,
$m_4=m_{\rho'}=\sqrt{2} \cdot 4 \pi f_{\pi}$. With these masses
the sum rules for the first moments, eq.\ (\ref{eq:1st}), are satisfied.
The actual, finite width resonances used in our analysis
follow this scheme, but with their widths fitted to the data
(see Appendix). The resulting
spectral functions $v_1(s)$ and $a_1(s)$ are shown in Fig. 1a,b.
By construction these model spectra satisfy the two WSR's
(\ref{eq:WSR1},\ref{eq:WSR2}).

The point to be emphasized is that the spectral strength
in the resonance region is well described by a pattern of
localized distributions, all of which integrate to even (for
vector channels) or odd (for axial-vector channels) multiples of 
$4 \pi^2 f_{\pi}^2$. Turning to finite
temperature $T$, we will now show that this statement survives to order $T^2$.

At finite $T$ the correlators are expressed in terms of their Gibbs
averages

\begin{equation} \label{eq:corrT}
\Pi^{\mu \nu}_{J}(q;T) = \frac{i \sum_n
\int d^4x\, e^{i q x}
\langle n | {\cal T} [J^{\mu} (x) J^{\nu} (0)] e^{-H/T} | n \rangle}
{\sum_n \langle n |e^{-H/T} | n \rangle}\, .
\end{equation}
The primary effect at low $T$ is a mixing of $V$ and $A$ modes through
their coupling to thermal pions. It was shown in ref.\
\cite{DeyEle} that, to leading order in $T^2$,

\begin{equation} \label{eq:corrTV} \nonumber
\Pi^{\mu \nu}_{V}(q;T) = (1- \varepsilon) 
\Pi^{\mu \nu}_{V}(q;T=0)+\varepsilon\Pi^{\mu \nu}_{A}(q;T=0)\, , \nonumber
\end{equation}

\begin{equation} \label{eq:corrTA}
\Pi^{\mu \nu}_{A}(q;T) = (1- \varepsilon) 
\Pi^{\mu \nu}_{A}(q;T=0)+ \varepsilon \Pi^{\mu \nu}_{V}(q;T=0)\, ,
\end{equation}
with $ \varepsilon\equiv T^2/(6f_{\pi}^2)$. To this order, poles in the
correlators remain at their positions; only their residues change.
Reducing eq.\ (\ref{eq:corrTA}) to its spin-0 and spin-1 components,
we derive for their thermal spectral functions
$v_{0,1}(s;T) = 2 \pi \mbox{Im} \Pi^{(0,1)}_V (s;T)$ and
$a_{0,1}(s;T) = 2 \pi \mbox{Im} \Pi^{(0,1)}_A (s;T)$:

\begin{eqnarray} 
v_0(s;T) &=& \varepsilon a_0 (s; T=0)\ ,\nonumber \\ 
a_0(s;T) &=& (1- \varepsilon) a_0 (s; T=0)\ ,\nonumber \\ 
v_1(s;T) &=& v_1(s;0)-\varepsilon [v_1(s;0)-a_1 (s; 0)]\ ,\nonumber \\ 
a_1(s;T) &=& a_1(s;0) + \varepsilon [v_1(s;0)-a_1 (s; 0)]\, .\label{eq:va1T}
\end{eqnarray}
Note that due to the coupling of the vector current to pions in the heat 
bath, the thermal $V$-correlator now receives an induced spin-0 (pion pole)
contribution. The thermal $V-A$ mixing is evident in eqs.\
(\ref{eq:va1T}). Moreover, the reduction of strengths in $a_0$, $v_1$
and $a_1$ by the common
factor $(1 - \varepsilon)$ is fully consistent with our previous
discussion:  to order $T^2$ the resonances
again contribute to the spectral integral as multiples of 
$f_{\pi}^2(T)= f_{\pi}^2(1-\frac{T^2}{6 f_{\pi}^2})$ \cite{GasLeuT}! Reversing 
the argument in the case of the pion pole, the $T$ dependence of
$f_\pi$ to this order was read off in Ref.\,\cite{DeyEle} by
defining $f_\pi^2(T)$ to be the residue at $T>0$.

For $s\ge s_c \simeq 5$ GeV$^2$ the continuum parts of the $V$
and $A$ spectra do not change with
$T$ to order $T^2$: since $v_1(s) = a_1(s)$ at $T=0$
for $s \ge s_c$, this is immediately evident from eq.\ (\ref{eq:va1T}).

Considering a given invariant 
(note that there are thermally induced pion pole terms in the vector channel), 
the FESR's for the lowest moments of the $V$ and $A$ spectral
distributions are not affected to order $T^2$. For example, consider

\begin{equation}
\left( \frac{4 q_{\mu} q_{\nu}}{3 q^4} - \frac{g_{\mu \nu}}{3 q^2}\right)
\Pi^{\mu \nu} (q;T)= \Pi^{(0)}(q^2;T)+\Pi^{(1)}(q^2;T)\, ,
\end{equation}
we have then

\begin{equation}
\int_0^{s_c} ds\, [v_0(s;T) +v_1(s;T)]= \int_0^{s_c} ds\, v_1(s;0)
- \varepsilon \int_0^{s_c} ds\, [ v_1(s;0)-a_1(s;0)-a_0(s;0)]\, .
\end{equation}
{Using (\ref{eq:pipole}) the
$T$ dependent part vanishes by virtue of the first Weinberg sum
rule (\ref{eq:WSR1}).} An analogous statement holds
for $a_0+a_1$. Furthermore, it is easily seen that

\begin{eqnarray}
\int_0^{s_c} ds\, s [v_0(s;T) +v_1(s;T)]&=& \int_0^{s_c} ds\, s v_1(s;0)\, ,\\
\int_0^{s_c} ds\, s [a_0(s;T) +a_1(s;T)]&=& \int_0^{s_c} ds\, 
s a_1(s;0)\nonumber\, ,
\end{eqnarray}
when using the second WSR, eq.\ (\ref{eq:WSR2}). {To
order $T^2$ the presence of the heat bath causes a mere redistribution
of spectral strength in both the $V$ and $A$ sectors but such that
the lowest moments of the spectral functions remain unchanged.} 
We illustrate these features by {an explicit calculation
at $T=140$ MeV with the results shown in Fig.\ 2.} It is amusing that already 
order $\varepsilon$ suggests the coincidence of vector and axial-vector
spectra at $T\simeq 160$ MeV ($\varepsilon = 1/2$) which is not far
from the critical temperature
$T_c \simeq 170$ MeV of the chiral phase transition 
determined by two-flavour lattice QCD \cite{Khan}. 

To summarize, based on the new experimental data we have shown that the 
resonant contributions to the spectral integrals in the $V$ and $A$ channels 
can be written as multiples of the square of the spontaneous
chiral symmetry breaking scale. Up to order $T^2$
this pattern carries over to the case of finite temperature.

\vspace{1cm}
{\large Acknowledgements:}\\
\vspace{.1cm}
One of us (R.H.) would like to thank 
Vladimir Eletsky for an explanation of his
own work and useful discussions.

\vspace{1cm}
{\large Appendix:}\\
\vspace{.1cm}

\noindent
To parametrize the resonances according to eqs.\ 
(\ref{eq:v1},\ref{eq:a0a1}) we 
use the following functional forms for $d_n$:

$$
d_1 \equiv d_{\pi} = \delta(s-m_{\pi}^2)\, ,
$$

$$
d_n = a_n \frac{\gamma_n^2(s)}{(s-m_n^2)^2 + \gamma_n^2(s)}\, 
$$

\noindent
where $\gamma_n(s) = [b_n + c_n (s - e_n)^2] \theta[b_n + c_n (s - e_n)^2]$,
with $n=2$ for $\rho$, $n=3$ for $a_1$ and $n=4$ for $\rho'$.
The parameter sets which reproduce the empirical spectra \cite{ALEPH,OPAL} are:

%\begin{H}{table}
\begin{center}
%\begin{tiny}
%\begin{tabular}{|c|ccccc|}
\begin{tabular}{cccccc}
%\hline
& $m_n/4 \pi f_{\pi}$& $a_n$ [GeV$^{-2}$]& $b_n$ [GeV$^2$]& $c_n$ [GeV$^{-2}$]
& $e_n$ [GeV$^2$]\\
\hline
$\rho$ & 0.63 & 3.7 & 0.143 & -0.07 & 1.2 \\
$a_1$ & 0.96 & 0.95 & 0.75 & -0.32 & 2.0\\
$\rho'$ & 1.42 & 0.502 & 1.3 & -0.3 & 2.6\\
\hline
\end{tabular}
%\end{tiny}
\end{center}
%\caption{}
%\end{table}
%
Note that the mass parameters $m_n$ deviate only {marginally} 
(by less than 10\% for $\rho$ and 3\% for $a_1$, $\rho'$)
from the expected ``large $N_c$'' pattern (see text), and the normalization
of the $d_n(s)$ required to reproduce the empirical data is equal
or close to one.

\begin{figure}
\begin{center}
\epsfig{file=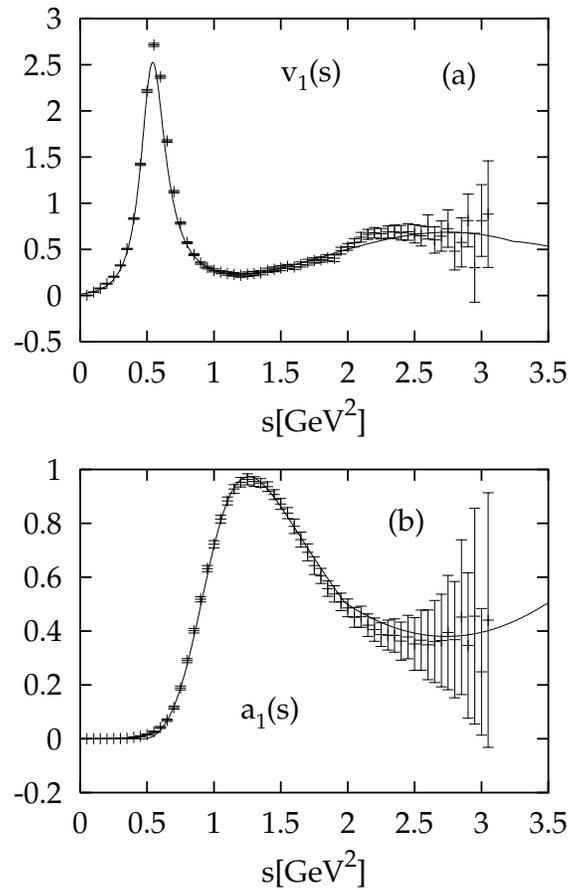,width=12cm,angle=-90}
\end{center}
\caption{Vector $(a)$ and axial-vector $(b)$ spectral
functions as given by the parametrization (\ref{eq:v1},\ref{eq:a0a1})
and Appendix, compared with ALEPH data \cite{ALEPH} (the comparison
with OPAL data \cite{OPAL} looks very similar).
}
\end{figure}

\begin{figure}
\begin{center}
\epsfig{file=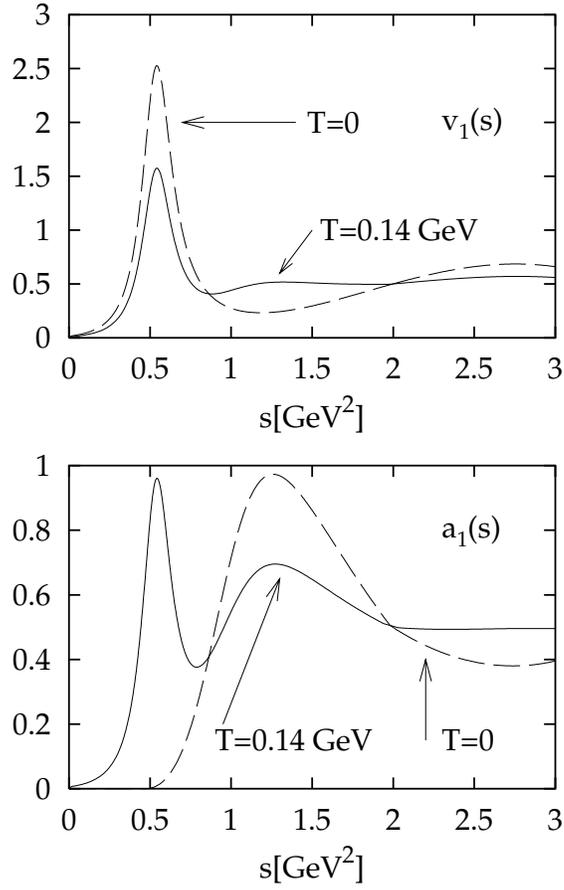,width=12cm,angle=-90}
\end{center}
\caption{Vector and axial-vector spectra calculated using eqs.\ 
(\ref{eq:v1},\ref{eq:a0a1}), at temperature $T=140$ MeV (solid lines)
as compared to $T=0$ (dashed lines).
}
\end{figure}

\end{document}